\documentclass[letterpaper,superscriptaddress,nobibnotes,footinbib,aps,prb,longbibliography,twocolumn,showpacs,floatfix]{revtex4-1}
\usepackage[utf8]{inputenc}
\usepackage[T1]{fontenc}
 
\usepackage{amsmath,amssymb,amsfonts}
\usepackage{amstext, mathrsfs, textcomp}
\usepackage{nicefrac}
\usepackage{multirow}

\usepackage{graphicx}
\usepackage{xcolor}

\usepackage[colorlinks=true, citecolor=blue, urlcolor=blue, linkcolor=blue]{hyperref}

\usepackage{lipsum}
\usepackage{changes}

\usepackage{notes2bib}
\bibnotesetup{
note-name
 = ,
use-sort-key = false
}

\graphicspath{{figures/}}

\newcommand{\TITLE}{Pushing the limits of optical information storage using deep learning}
\newcommand{\comment}[1]{}

\hypersetup{
	pdftitle = {\TITLE},
	pdfsubject = {},
	pdfauthor = {Peter R. Wiecha et al.},
	pdfkeywords = {},
	pdfcreator = {LaTeX},
	pdfproducer = {LaTeX with hyperref and thumbpdf}
}

\begin{document}

%%%%%%%%%%%%%%%%%%%%%%%%%%%%%%%%%%%%%%%%%%%%%%%%%%%%%%%%%%%%%%%%%%%%%%%%%
%% REVTEX
%%%%%%%%%%%%%%%%%%%%%%%%%%%%%%%%%%%%%%%%%%%%%%%%%%%%%%%%%%%%%%%%%%%%%%%%%
\title{\TITLE}
% \subtitle{}

\author{\firstname{Peter R.} \surname{Wiecha}}
\email[e-mail~: ]{peter.wiecha@cemes.fr}
\affiliation{CEMES, Universit\'e de Toulouse, CNRS, Toulouse, France}

\author{Aur\'elie Lecestre}  % EBL litho
\affiliation{LAAS, Universit\'e de Toulouse, CNRS, INP, Toulouse, France}

\author{Nicolas Mallet}  % EBL litho
\affiliation{LAAS, Universit\'e de Toulouse, CNRS, INP, Toulouse, France}

\author{Guilhem Larrieu}  % EBL litho
\affiliation{LAAS, Universit\'e de Toulouse, CNRS, INP, Toulouse, France}

%%%%%%%%%%%%%%%%%%%%%%%%%%%%%%%%%%%%%%%%%%%%%%%%%%%%%%%%%%%%%%%%%%%%%%%%%
%%%%%%%%%%%%%%%%%%%%%%%%%%%%%%%%%%%%%%%%%%%%%%%%%%%%%%%%%%%%%%%%%%%%%%%%%

\begin{abstract}
Diffraction drastically limits the bit density in optical data storage. To increase the storage density, alternative strategies involving supplementary recording dimensions and robust read-out schemes must be explored.
Here, we propose to encode multiple bits of information in the geometry of subwavelength dielectric nanostructures.
A crucial problem in high-density information storage concepts is the robustness of the information readout with respect to fabrication errors and experimental noise.
Using a machine-learning based approach in which the scattering spectra are analyzed by an artificial neural network, we achieve quasi error free read-out of sequences of up to 9 bit, encoded in top-down fabricated silicon nanostructures.
We demonstrate that probing few wavelengths instead of the entire spectrum is sufficient for robust information retrieval and that the readout can be further simplified, exploiting the RGB values from microscopy images.
Our work paves the way towards high-density optical information storage using planar silicon nanostructures, compatible with mass-production ready CMOS technology.
\end{abstract}

\maketitle

%% ---------------------------------------------- SECTION: Introduction
% \section{Introduction}

Optical information storage promises perennial longevity, high information densities and low energy consumption compared to magnetic storage media.\cite{zhang_seemingly_2014, gu_optical_2014}
The compact disc (CD) and its successors, the DVD and the blue-ray disc, broadly established optical storage in our society.\cite{satoh_key_1998, borg_phase-change_2001}
Those media are based on storing a single binary digit per diffraction limited area (``zero'' or ``one'').
Several concepts have been proposed to increase the information density in optical storage. 
Examples are schemes exploiting polarization-sensitive digits,\cite{zeng_polarization-based_2014}
near-field optical recording,\cite{tominaga_approach_1998} the use of fluorescent dyes \cite{mottaghi_thousand-fold_2013}
or three-dimensional approaches like two-photon point-excitation\cite{strickler_three-dimensional_1991}.
Yet, all these alternatives suffer from major drawbacks. 
Either they are hardly superior to commercial planar solutions (polarization-sensitive patterns) or they require very complex storage media (fluorescence) or sophisticated read-out schemes (near-field recording, two-photon point-excitation).
The most promising alternative seemed to be holographic memory, which was proposed in the early 1960ies and makes use of the volume of the storage medium. 
To date, however, there is still no commercial product available, despite several announcements in the past 20 years.\cite{heerden_theory_1963, psaltis_holographic_1998}

In the last decades, photonic nanostructures emerged as powerful instruments to control light at the nanometer scale.\cite{girard_near_2005, novotny_principles_2006}
Localized surface plasmons (LSP) in metal nanoparticles\cite{maier_plasmonics_2010} or Mie-type resonances in high-index dielectric structures\cite{kuznetsov_optically_2016} cover the entire visible spectrum and can be tuned by designing appropriate geometric features.\cite{cao_tuning_2010, wiecha_evolutionary_2017}
Furthermore, the high scattering efficiencies of photonic nanostructures render single-particle spectroscopy relatively easy.
In consequence, the idea has been raised to encode information in the rich scattering spectra of plasmonic nanostructures, denser than a single data bit.\cite{mansuripur_plasmonic_2009, chen_manipulation_2011, cui_plasmonic_2014, el-rabiaey_novel_2016, zijlstra_five-dimensional_2009}
The information density might be even further increased by addressing layer-wise arranged nanostructures via the focal depth\cite{taylor_detuned_2012} or by the polarization of the probe light\cite{taylor_electron-beam_2014}.
A key problem of such approach is the availability and accuracy of read-out schemes.\cite{chen_manipulation_2011, li_multifocal_2015}
The main difficulty lies in the fact that different nanostructure geometries can lead to quite similar optical responses, which need to be unambiguously identified during the information retrieval.\cite{chen_manipulation_2011, liu_training_2018}
This problem is further complicated by structural defects of the fabricated particles and by the noise generated during the optical detection.
Further drawbacks are associated with the metallic character of plasmonic nanostructures. 
One problem is the limited scalability of the production. 
Another inconvenience with gold (the most common plasmonic material) is the limitation to wavelengths above the interband transitions, hence larger than \(\approx 520\,\)nm. 
Shorter wavelengths cannot be used to encode information, which effectively reduces the attainable information density.

%%%%%%%%%%%%%%%%%%%%%%%%%%%%%%%%%%%%%%%%%%%%%%%%%%%%%%%%%%%
\begin{figure*}[t!]
	\begin{center}
		\includegraphics[page=1]{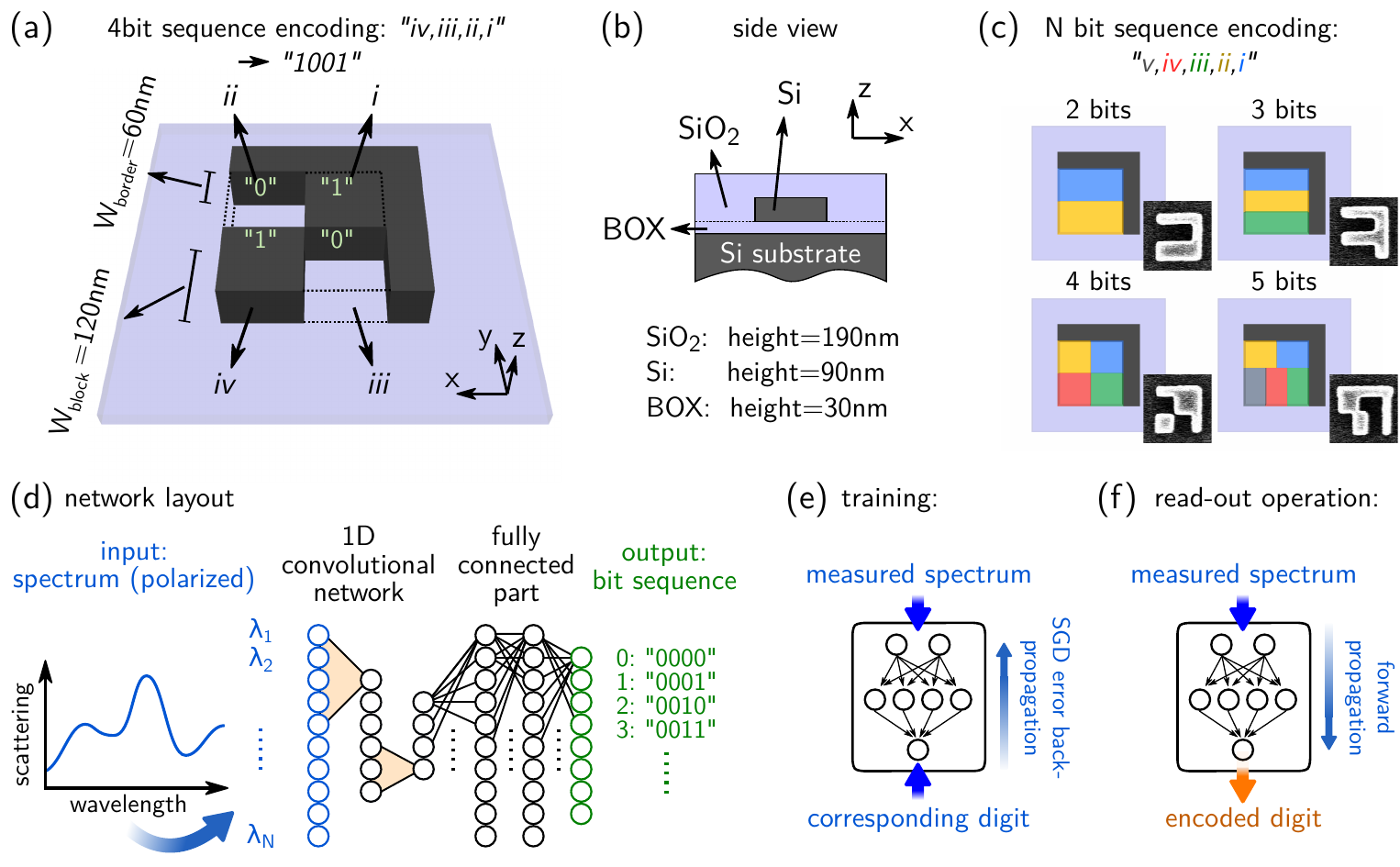}
		\caption{%
			\textbf{Sketch of the structure geometry and the 1D convolutional artificial neural network.}
			(a) Illustration of the ``4 bit'' nanostructure geometry. 
			The digital information is encoded in the four silicon blocks (block: ``1'', no block: ``0'').
			The structure corresponds to the 4 bit digit ``1001'' (decimal ``9'').
			The L-shaped sidewall is necessary to distinguish symmetric arrangements via polarized optical spectroscopy.
			(b) side-view of the fabricated structure layout.
			(c) Sketches of the different geometry models used to experimentally encode sequences of 2, 3, 4 or 5 bits in each nanostructure. 
			In all cases, the height is \(90\,\)nm and the L-shaped wall is \(60\,\)nm wide. 
			In the 2 bit geometry, the blocks are \(120\times 240\,\)nm\(^2\) large. 
			The 3 bit geometry uses blocks of \(120\times 240\,\)nm\(^2\). For the 4 bit encoding, the blocks occupy areas of \(120\times 120\,\)nm\(^2\). 
			In the 5 bit geometry, the two upper blocks measure \(150\times 100\,\)nm\(^2\) and the three lower ones have a size of \(100\times 150\,\)nm\(^2\).
			Insets of selected SEM images show areas of \(450\times 450\,\)nm\(^2\).
			(d) sketch of the 1D convolutional neural network used for the classification task.
			(e) training stage of the artificial neural network: measured spectra and corresponding digital information are fed into the network. The error is back-propagated using a variant of the stochastic gradient descent (SDG) algorithm.
			(f) the trained network is capable to retrieve the digital information encoded in the structures via their spectra.
		}\label{fig:fig1}
	\end{center}
\end{figure*}
%%%%%%%%%%%%%%%%%%%%%%%%%%%%%%%%%%%%%%%%%%%%%%%%%%%%%%%%%%%

To overcome all of these limitations, we develop here a scheme for digital information encoding, based on silicon nanostructures.
Owing to its high refractive index with low imaginary part, silicon nanostructures support low-loss optical resonances, tunable over the entire visible spectral range.\cite{albella_low-loss_2013, kuznetsov_optically_2016, wiecha_evolutionary_2017}
In addition, silicon has great technological advantages,
first of all the mass-production ready, high-precision complementary metal-oxide-semiconductor (CMOS) based processing technology, and its low cost and durability.
To reliably retrieve the stored information, we propose a machine learning (ML) based approach, in which the scattering spectra are analyzed by an artificial neural network (ANN).
ANNs are computational schemes that can be ``trained'' to efficiently solve problems, hard for classical computer arithmetics.\cite{nielsen_neural_2015, goodfellow_deep_2016}
%Furthermore, such networks are capable of a certain degree of abstraction, in other words they can develop the ability to correctly classify noisy data, never seen during the training.
ANNs are used in many every-day applications ranging from spelling correction, sentence completion and image recognition in modern smartphones to medical image interpretation.\cite{szegedy_inception-v4_2016, mamoshina_applications_2016}
Besides a few recent examples, ML is being scarcely applied on problems in nano-optics. 
In one work, robust data read-out from holographic memory was realized using convolutional ANNs.\cite{shimobaba_convolutional_2017}
The potential of ANNs has recently been demonstrated also in classification and inverse design of nanoparticles.\cite{jo_holographic_2017, malkiel_plasmonic_2018, peurifoy_nanophotonic_2018}
%Our structures are fabricated on commercial SOI substrate and each sub-diffraction limit small particle encodes 4 bit of information in its geometry.
We train the digital-information retrieval neural network, on the measured scattering spectra of several hundred fabricated copies of each nanostructure corresponding to a binary sequence.
%We then train an ANN on the individually measured scattering spectra.
%Once trained on our dataset, the ANN is capable to classify formerly unseen nano-structures by their scattering spectra. 
On all our experimental datasets encoding up to 9 bit of information, effectively going beyond the data density of the blue-ray disc, the trained ANNs yield a quasi error-free read-out.
We demonstrate furthermore that an accurate retrieval can be achieved using the scattering intensity at only a small number of discrete wavelengths or even simply the RGB (red-green-blue) color values from darkfield (DF) microscopy images.
The latter approach allows in principle a massively parallel read-out of the stored data.

%%%%%%%%%%%%%%%%%%%%%%%%%%%%%%%%%%%%%%%%%%%%%%%%%%%%%%%%%%%
\begin{figure*}[t!]
	\begin{center}
		\includegraphics[page=1]{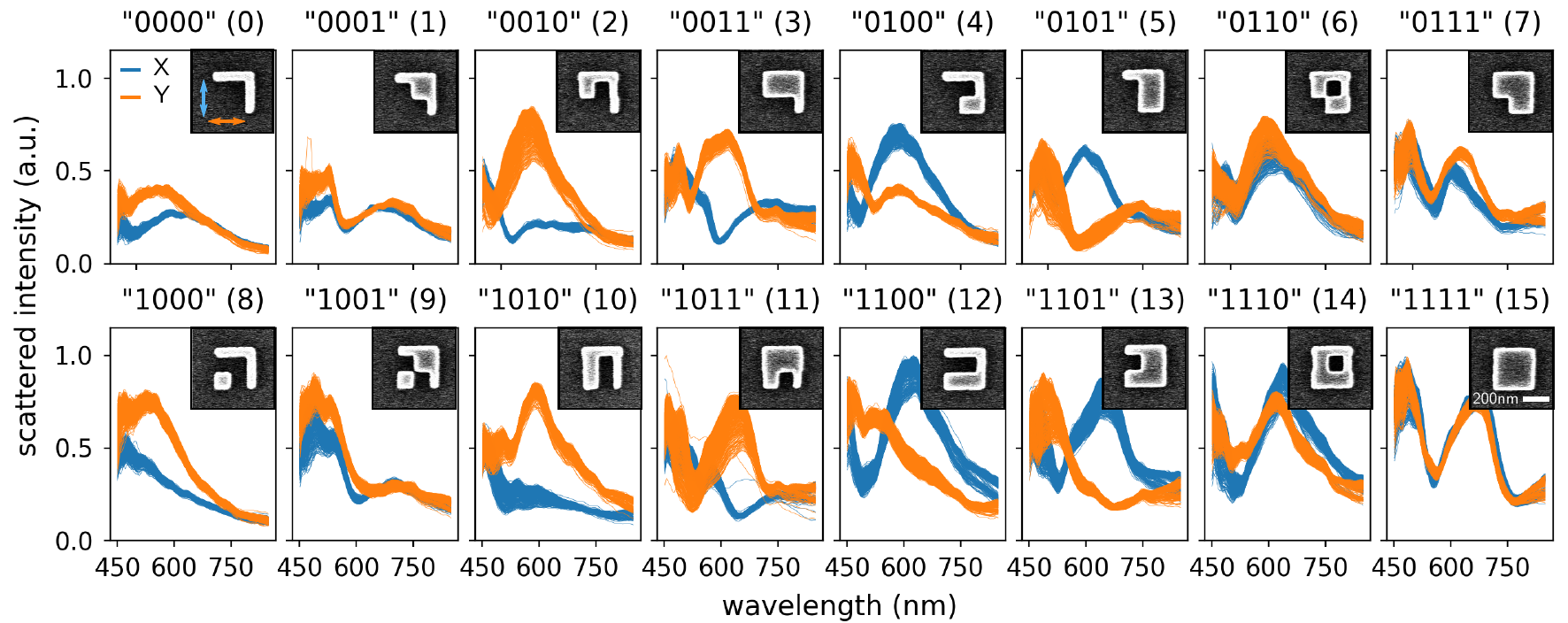}
		\caption{
			\textbf{Experimental darkfield spectra training data-set for 4 bits.} Our data comprises measurements from 625 copies for each of the 16 ``4 bit'' geometries (this makes a total of \(625\times 16\times 2 = 20000\) acquired spectra).
			The spectra are superposed above each other.
			Blue lines: DF scattering for \(X\)-polarized light, orange lines: \(Y\)-polarization.
			Insets show SEM images of one representative copy of the respective nanostructure, the areas are \(600 \times 600\,\)nm\(^2\) large, the scalebar in ``1111'' is \(200\,\)nm. 
			The not shown 2 bit dataset consists of 5000, the case of 3 bits comprises 10000 and the 5 bit set contains 40000 scattering spectra (all shown together with SEM images in the {\color{blue}supporting informations}, (Figs.~S4-S11\comment{CHECK}), where also a separate 4 bit dataset and simulations are shown in order to demonstrate the reproducibility (Figs.~S12-S14\comment{CHECK}).
		}\label{fig:fig2}
	\end{center}
\end{figure*}
%%%%%%%%%%%%%%%%%%%%%%%%%%%%%%%%%%%%%%%%%%%%%%%%%%%%%%%%%%%

\section{Silicon nanostructures for digital information encoding}

As illustrated in figure \ref{fig:fig1}a-b, we use a planar array-like geometry to encode several bits of information in a single silicon nanostructure.
If a certain position in the 2D array contains a silicon block, the according bit is set to ``1'', otherwise it is ``0''.
In order to unambiguously distinguish symmetric or rotational arrangements (for example 4 bit ``0010'' and ``0100''), an L-shaped silicon frame is added, surrounding two sides of the structure.
In this way, under linearly polarized illumination each binary number yields a unique spectral response.

For a first demonstration we fabricate nanostructures encoding between 2 and 5 bits of information each, as illustrated in figure~\ref{fig:fig1}c, using electron-beam lithography (ebeam) and subsequent dry-etching of commercial silicon-on-insulator (SOI) substrates with a silicon overlayer of \(H=90\,\)nm height.
Subsequent to the etching, the structures are covered by a protective SiO\(_2\) layer of \(190\,\)nm height (see figure~\ref{fig:fig1}b). 
For more details on the fabrication process, see the Methods section.

We fabricate \(25\times 25 = 625\) copies of each geometry. 
Using an automated setup with an \(XY\) piezo stage, we measure the linearly \(X\) and \(Y\) polarized DF spectra of each copy of the structures.
All acquired spectra for the 4 bit case together with representative scanning electron microscopy (SEM) images are shown in figure~\ref{fig:fig2}, superposed and grouped according to the 16 digital numbers.
The spectra for the 2, 3 and 5 bit datasets can be found in the {\color{blue}supporting informations (SI)}, Figs.~S4-S11\comment{CHECK}.
We note that the spectra of symmetric structures are not completely identical for crossed polarizations (see \textit{e.g.} ``0000'').
We attribute this observation to the ebeam being horizontally rasterscanned in the fabrication process, leading to small anisotropies (about \(5-10\%\)) between horizontal and vertical features (see SEM images in SI Figs.~S1-S3 and~S21\comment{CHECK}). 
This can in principal be corrected in an automated fashion during the mask design, as demonstrated in the SI, Figs.~S27-S28\comment{CHECK}.
On the other hand, such asymmetries can be even advantageous for our purpose, because they increase the ``uniqueness'' of the individual geometries and their scattering spectra.

%%%%%%%%%%%%%%%%%%%%%%%%%%%%%%%%%%%%%%%%%%%%%%%%%%%%%%%%%%%
\begin{figure*}[t!]
	\begin{center}
		\includegraphics[page=1]{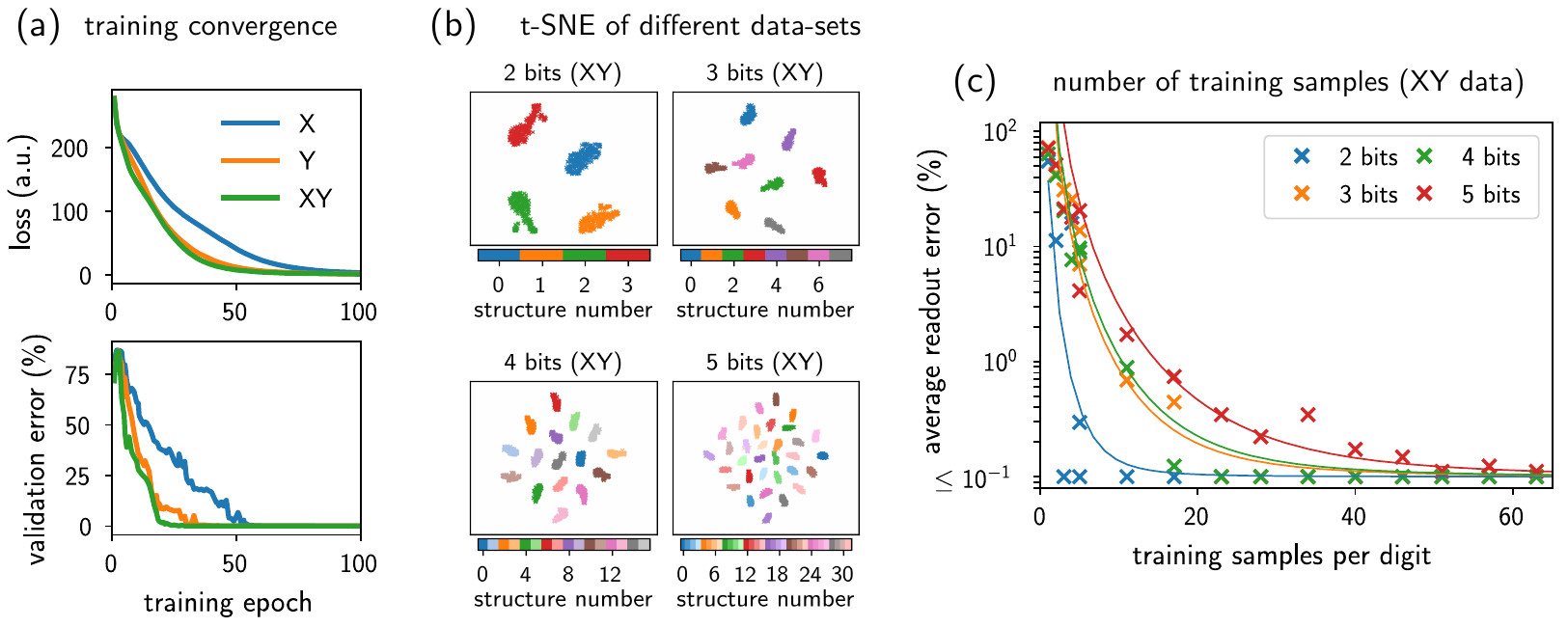}
		\caption{
			\textbf{Training convergence and read-out accuracy of the ANN trained on the full scattering spectra.}
			(a) convergence of the training at the example of 4 bit encoding and spectra of 300 copies per digit for training. 
			Loss (categorical cross-entropy, top) and validation error rate (bottom) as function of the training epoch for only \(X\) (blue), only \(Y\) polarized spectra (orange) and both polarizations (green) as training data.
			(b) t-SNE\cite{van_der_maaten_visualizing_2008} visualization (see also text) of the ``XY'' polarization training sets for 2-5 bits encoded per structure.
			(c) read-out error as function of the number of spectra per structure used for the training of the network.
			All cases of information density per structure yield quasi error-free read-out using sufficiently large training sets of at most around 60 samples per geometry (5 bit case, red crosses).
			As one would expect, the lower bit densities (blue: 2 bit, orange: 3 bit, green: 4 bit) require even less training samples for an accurate information retrieval.
			Solid lines are guides to the eye, proportional to \(N^{-3}\), which qualitatively describe the convergence of read-out as function of training set size \(N\). 
%			The dashed purple line is a qualitative extrapolation of the parameter \(A\), based on the experimental results for 2-5 bits (see also {\color{blue}SI, {\color{red}Figs~SXXXX}}).
		}\label{fig:fig3}
	\end{center}
\end{figure*}
%%%%%%%%%%%%%%%%%%%%%%%%%%%%%%%%%%%%%%%%%%%%%%%%%%%%%%%%%%%

\section{Machine learning based digital information read-out}

\subsection{Read-out using scattering spectrum}

Our goal is to read the information, encoded in the geometries of the silicon structures via a far-field optical measurement scheme.
The optical scattering spectrum is a promising physical quantity to differentiate between the different structures, in other words to retrieve the bit-sequences they represent.
Here, we propose a machine-learning approach to the problem.
We train an ANN using sub-sets of the acquired spectra.
Subsequently, we evaluate the accuracy of the read-out by testing the trained ANN with spectra not used for training.

\subsubsection{Network architecture}

We use a one-dimensional convolutional neural network (ConvNet), followed by a fully connected network, as depicted in figure \ref{fig:fig1}d -- an architecture with particular strength at pattern-recognition tasks.\cite{goodfellow_deep_2016}
The spectra are fed in the network input layer which consists of one or two parallel channels, depending on whether a single or both polarization cases are used. 
At the ``softmax'' output layer, each bit sequence is attributed to one neuron.
Details on the network and training parameters as well as on the preprocessing of the data can be found in the Methods section.
The network loss and the error rate on the validation set are shown in figure~\ref{fig:fig3}a for the first 100 epochs\footnote{An ``epoch'' is a ML term, signifying one full training iteration. In each epoch, the full training dataset is randomly shuffled and used in its totality to optimize the network parameters.} of training on the 4 bit dataset.

\subsubsection{Results}

The read-out scheme is illustrated in Fig.~\ref{fig:fig1}f: The scattering spectra of the binary structures are fed into the trained ANN and forward propagated through the network. 
The output neuron with the highest activation indicates the encoded bit sequence.
In all cases (2-5 bits encoded per structure; using either \(X\) or \(Y\) or both (``XY'') polarizations), the trained ANN yields quasi error free read-out accuracy.
In the cases of 4 and 5 bits, one single spectrum of the test-data was incorrectly interpreted (corresponding to 0.023\% and 0.011\% for the 4400 (4 bit), respectively 8800 (5 bit) test-structures).

We analyzed the datasets using the ``t-SNE'' dimensionality reduction, in order to estimate the distinguishability between different geometries, as well as the variance in the spectra from copies of identical structures.\cite{van_der_maaten_visualizing_2008}
In a t-SNE plot, well separated scatter points correspond to unambiguously differentiable entities in the dataset.
Nearby and overlapping points on the other hand correspond to very similar data.
The results for the ``XY'' datasets are shown in figure \ref{fig:fig3}b, the t-SNE plots for the ``X'' and ``Y'' datasets can be found in the {\color{blue}SI}, Fig.~S15\comment{CHECK}.
Each color corresponds to one of the binary structures, each dot represents the measurement from one specific copy.
All datasets are characterized by a very good separation of the different spectra in the t-SNE plots, which explains why the ANN can retrieve the binary information with almost no errors.
Figure~\ref{fig:fig3}c shows the readout error as function of the training samples.
As intuitively expected, the required number of training spectra for error-free operation increases with the complexity of the geometrical model. 
On the other hand, we observed that the main source for dispersion in the spectra are the measurement conditions. The shape of the spectra is particularly sensitive to the position of the confocal hole with respect to the nanostructure. Hence, the spectral dispersion could be easily reduced by optimizing the stability of the acquisition scheme.
We also want to emphasize that our conditions are still perfectly sufficient for a quasi error-free operation in all considered cases.

%%%%%%%%%%%%%%%%%%%%%%%%%%%%%%%%%%%%%%%%%%%%%%%%%%%%%%%%%%%
\begin{figure*}[t!]
	\begin{center}
		\includegraphics[page=1]{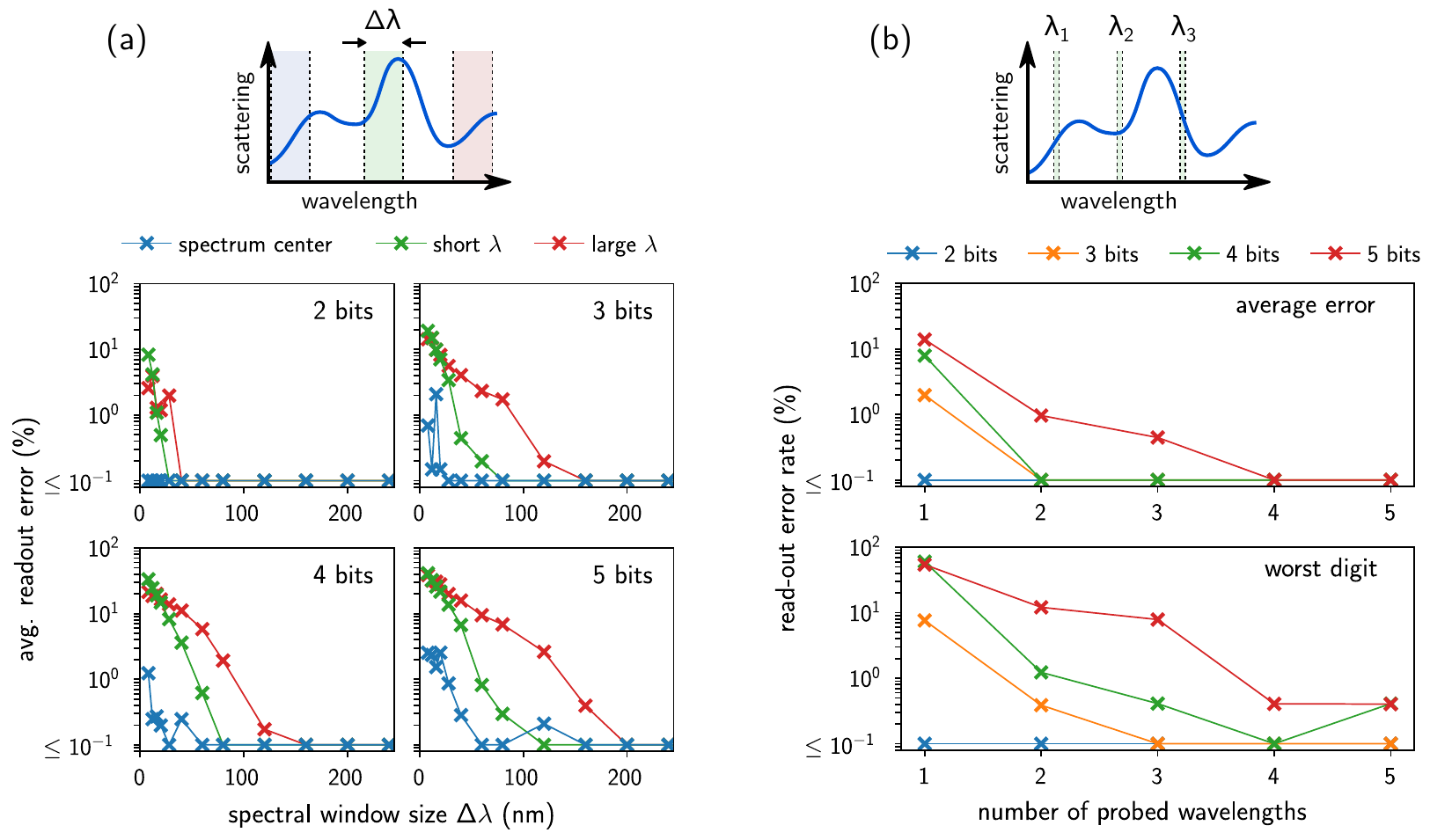}
		\caption{
			\textbf{Accuracy of network trained on reduced spectral information.}
			(a) Training using scattering from spectral window of reduced width for the 2, 3, 4 and 5 bit datasets (left top to right bottom plot).
			The scattering intensity was taken from a window either at the short wavelength side (blue lines), in the center (green lines) or from the red edge of the spectra.
			(b) Average (top plot) and worst-digit (bottom pot) read-out accuracy using a network trained on the scattering intensity of \(X\)- and \(Y\)-polarized light at a discrete number of wavelengths.
            The explicit positions of \(\lambda_i\) for the different cases are given in the Methods section.
			t-SNE plots for all reduced spectral information datasets can be found in the {\color{blue}SI}, Figs.~S15.\comment{CHECK}
		}\label{fig:fig4}
	\end{center}
\end{figure*}
%%%%%%%%%%%%%%%%%%%%%%%%%%%%%%%%%%%%%%%%%%%%%%%%%%%%%%%%%%%

To assess the amount of optical data required for an accurate readout, we train and test ANNs using reduced spectral information.
Figure~\ref{fig:fig4}a shows the error rate as function of the spectral window width \(\Delta \lambda\) used for training and retrieval (on the ``XY'' datasets).
We evaluated three different positions of the spectral window: 
Either the scattering is taken at the long wavelength end of the spectra (\(\lambda \leq 850\,\)nm, red lines), at short wavelengths (\(\lambda \geq 450\,\)nm, blue lines), or it is centered around \(650\,\)nm (green lines).
While the short and intermediate wavelengths always yield error free readout for spectral windows as small as \(\lesssim 100\,\)nm, we observe that using the red part of the spectra requires a larger spectral window of up to \(\approx 200\,\)nm for high accuracy.
We conclude that the red part of the spectra contains the least amount of information, not sufficient to unambiguously distinguish between the binary sequences.
This is a direct consequence of the photon energy being inverse proportional to the wavelength.
Furthermore, the used geometries have no resonances above \(\approx 750\,\)nm.
Using larger or higher structures could increase the information density in the red by shifting resonances to longer wavelengths.
In a second step, we train a fully connected ANN (see also Methods) using the scattering intensity only at a low number of discrete wavelengths, as shown in figure~\ref{fig:fig4}b.
In all cases, probing three wavelengths is sufficient to obtain a readout accuracy of \(>99\,\)\%. While the error rates on the worst bit-sequence are still in the order of \(10\,\)\%, the worst digit error can be reduced below \(0.5\,\)\% by probing at 4 or 5 wavelengths.

%%%%%%%%%%%%%%%%%%%%%%%%%%%%%%%%%%%%%%%%%%%%%%%%%%%%%%%%%%%
\begin{figure*}[t!]
	\begin{center}
		\includegraphics[page=1]{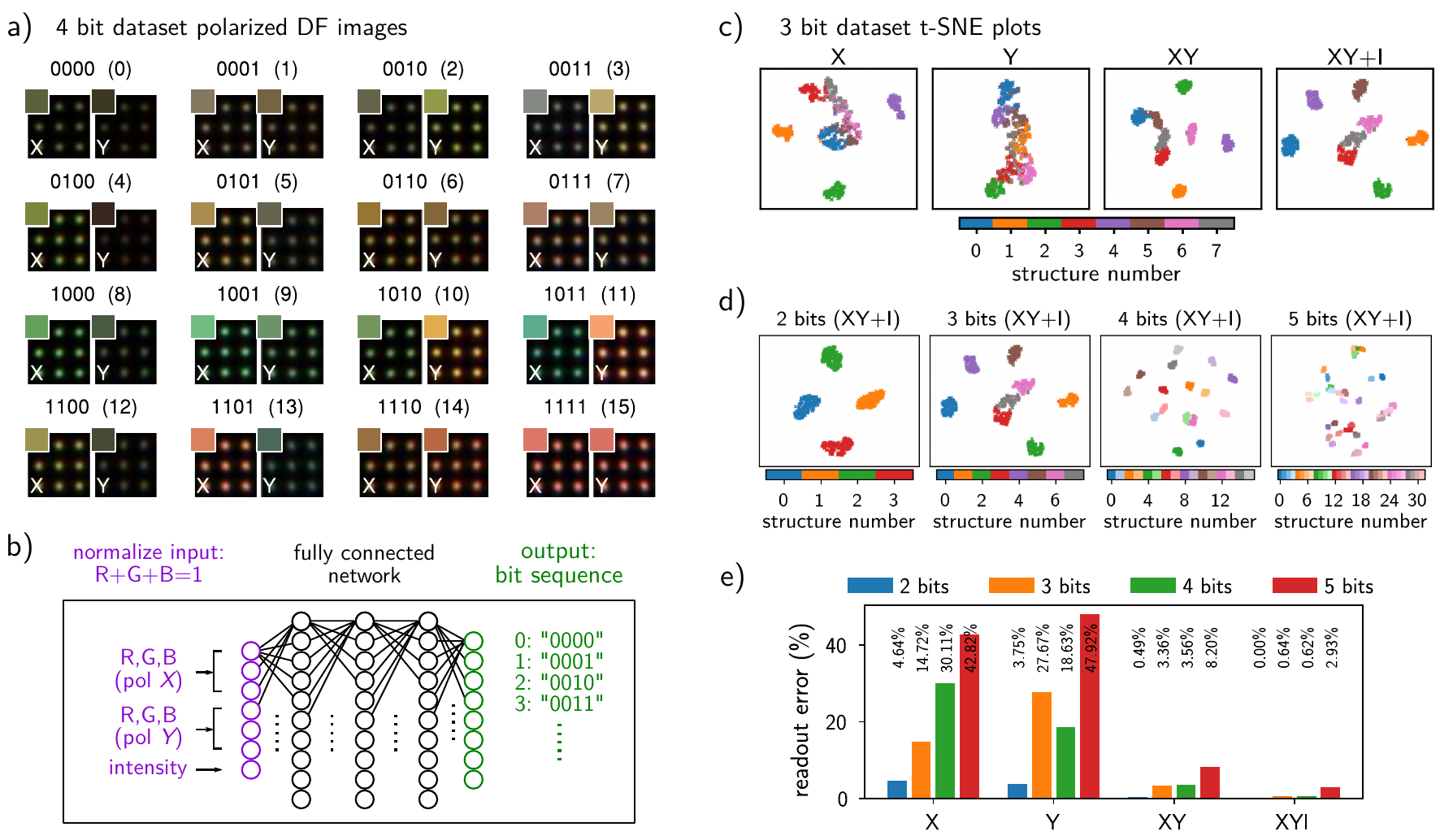}
		\caption{
			\textbf{Neural-network based data read-out via the RGB color values.}
			(a) polarization filtered dark-field color images of representative \(3\times 3\) arrays of the ``4 bit'' digit structures.
			Left: \(X\)-polarization, right: \(Y\)-polarization.
			The DF images show areas of \(7.5\times 7.5\,\)\textmu m\(^2\).
			The insets show the average RGB color of the \(3\times 3\) structures.
			(b) scheme of the fully connected artificial neural network used for the RGB classification task.
			(c) t-SNE\cite{van_der_maaten_visualizing_2008} visualization of the ``3 bit'' training sets for only \(X\), only \(Y\), \(X\) plus \(Y\) or \(X\), \(Y\) and scattered intensity \(I\) (from left to right).
			Only if using simultaneously both polarizations (``XY'' and ``XY+I''), the different bit sequences show a clear separation in the t-SNE plots.
			(d) t-SNE plots for the XY+I cases of the 2, 3, 4 and 5 bit training sets (from left to right).
			(e) information retrieval accuracy of the network, trained on the different data-sets consisting of only the \(X\)-filtered, only the \(Y\)-filtered, \(X\)+\(Y\) filtered  and \(XY\) + the scattered intensities \(I\).
			t-SNE plots for all data-sets can be found in the {\color{blue}SI}, Fig.~S15.\comment{CHECK}
		}\label{fig:fig5}
	\end{center}
\end{figure*}
%%%%%%%%%%%%%%%%%%%%%%%%%%%%%%%%%%%%%%%%%%%%%%%%%%%%%%%%%%%

\subsection{Parallel read-out using RGB color from microscope images}

Obviously, reduced spectral information is sufficient for accurate information retrieval.
Therefore, we will study if the bit sequences can be recovered also using a simpler and faster data acquisition scheme, namely the scattered RGB color obtained from conventional DF microscopy images.

\subsubsection{Training data and network architecture}

Figure~\ref{fig:fig5}a shows DF images of \(3\times 3\) copies for each geometry and both polarizations at the example of the 4 bit geometry. 
The average RGB color is shown in the upper left corner of each plot.
For training we use the average RGB values from the scattered light of each individual structure in the polarization filtered DF image.
We then normalize the RGB values to \(R+G+B=1\) and separately store the scattered intensity.
We create four training datasets: Three sets with only the normalized RGB information. 
One for \(X\), and one for \(Y\) polarized data, as well as a third dataset combining both polarizations (``XY''). 
The forth set contains the XY data and additionally their intensity values (``XY+I'').
We use an entirely fully connected network architecture, as depicted in Fig.~\ref{fig:fig5}b. 
Technical details are given in the Methods section.

\subsubsection{Results}

Figures~\ref{fig:fig5}c and~d respectively show ``t-SNE'' plots for the different 3 bit datasets and for the ``XY'' data of the 2, 3, 4 and 5 bit geometries.
The partial mixing of scatter-points in the t-SNE plots of only \(X\) or only \(Y\) polarized RGB values suggests that these data are not sufficient for accurate identification of the structures.
This conjecture is confirmed by the insufficient readout performance of the corresponding ANNs, as shown in figure~\ref{fig:fig5}e. 
The error rates of several individual binary numbers are even well above 50\% (see also {\color{blue}SI}, Fig.~S16\comment{CHECK}).

Using the datasets combining both polarizations, on the other hand, the situation drastically improves. 
The t-SNE plots (Fig~\ref{fig:fig5}d) now show a clear separation of the different binary numbers. 
The average error rate drops significantly below 10\,\% in the ``XY'' case, and can be reduced further below 1\% (2-4 bits) and below 3\% (5 bits) when the brightness values are considered as well.
In the 4 bit ``XY+I'' case for instance, the largest error rate is observed for digits ``0101'' and ``1100'' which scatter light in a resembling tone and brightness (see also {\color{blue}SI} Figs~S16).\comment{CHECK}
The similarity can also be observed in the t-SNE plot (figure~\ref{fig:fig5}d), where the light green and pink dots (structures ``5'' and ``12'') are very close and partly touching.
Through such an analysis of the t-SNE plots, the most problematic digits can also be identified in the other datasets (e.g. ``6'', ``8'', ``29'' and ``31'' in the 5 bit case).
By designing nanostructures with a more significant difference in the scattered colors (e.g. using modern inverse problem techniques\cite{wiecha_evolutionary_2017, feichtner_evolutionary_2012}), this limitation could be easily overcome and the error rate further decreased.
In Figs.~S17-S18\comment{CHECK} of the {\color{blue}SI}, the activations of the neurons of the softmax output layer of the ``RGB'' network are shown for the whole validation sets of the 4 bit datasets ``X'' and ``XY+I''.

The RGB color information allows to simultaneously capture many thousands of structures within a single measurement of a large-area image.
In other words it supports a massively parallel read-out of the information (see also Methods).
Cheap, smartphone-based DF microscopy for the RGB read-out might become feasible in the near future. 
Indeed, smartphone-based microscopy is subject of current research and has undergone tremendous progress in the recent past.\cite{rivenson_deep_2018, orth_dual-mode_2018, wei_plasmonics_2017}
A further interesting route for improvement are specifically designed bright-field color scatterers\cite{flauraud_silicon_2017} for information encoding, in order to avoid the necessity of the complex DF illumination scheme.
Finally, a cheap multi-laser approach, similar to a blue-ray disc reader with several lasers could be used to capture the scattering intensity simultaneously at several wavelengths.

%%%%%%%%%%%%%%%%%%%%%%%%%%%%%%%%%%%%%%%%%%%%%%%%%%%%%%%%%%%
\begin{figure*}[t!]
	\begin{center}
		\includegraphics[page=1]{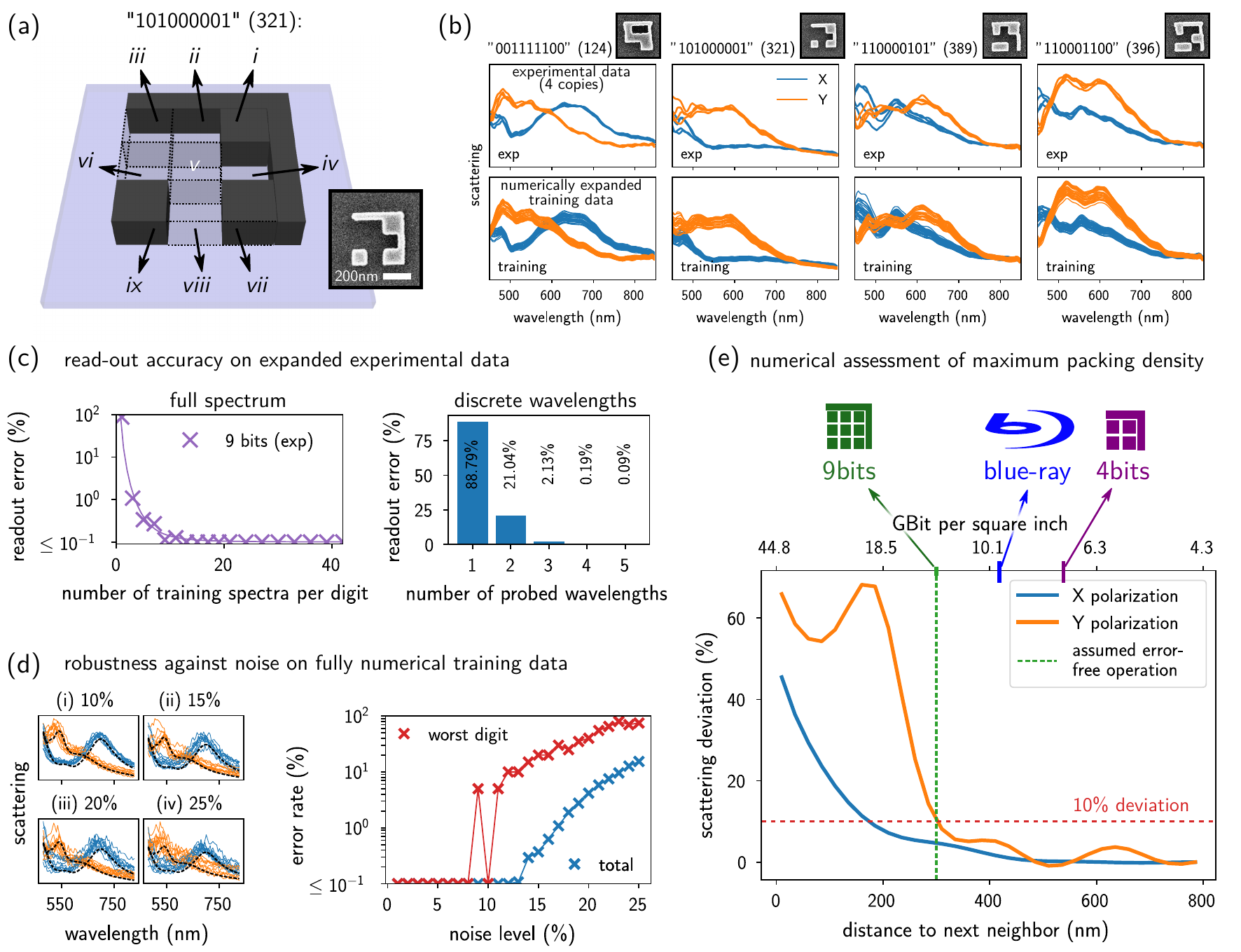}
		\caption{
			\textbf{9 bit per nanostructure information encoding.}
			(a) geometry of a silicon nanostructure encoding \(3\times 3 = 9\) bits (512 possible combinations). 
			Each silicon block occupies an area of \(105\times 105\,\)nm\(^2\). The L-shaped sidewall is \(45\,\)nm wide. The height is \(90\,\)nm.
			An SEM image of a fabricated structure is given in the inset, where the scalebar is 200\,nm.
			The shown example represents the decimal number ``321''.
			(b) selected examples, illustrating the training data generation by a numerical expansion of the experimental spectra.
			SEM images show areas of \(550\times 550\,\)nm\(^2\).
			We fabricated 4 copies of each possible 9 bit nanostructure. 
			Via random superposition of the experimental scattering spectra of these 4 copies, we generate a large set of spectra, allowing us to train and test the performance of the binary information readout ANN (see also {\color{blue}SI} Fig.~S23\comment{CHECK} for more details).
			(c) accuracy of the ANN trained on the experimental 9 bit data using the full spectra (left, the solid line is a guide to the eye) or scattering intensities at a limited number of discrete wavelengths (right). In both cases, \(X\) and \(Y\) polarized data is used simultaneously.
			(d) evaluation of the robustness of 9 bit read-out with respect to noise on fully numerical data.
			We use simulated scattering spectra for all 9 bit geometries with different amounts of random noise (see {\color{blue}SI} Fig.~S33\comment{CHECK} for details on the training data generation).
			Example spectra for noise levels from (i) 10\% to (iv) 25\% are shown at the example of structure ``001011010'' (decimal 90).
			The plot on the left shows the readout error rate on the numerical data as function of noise level (training on the full spectra of both polarizations).
			(e) scattering deviation relative to the isolated structure as function of the distance between two digit encoding structures for focused illumination. 
			For the estimation of the feasible information density, 10\% deviation from the unperturbed spectrum are assumed to be tolerable (red dashed horizontal line).
			This leads to an information density (green indicator) around \(40\)\% higher than the blue-ray disc (blue indicator). 
			The 4 bit structures yield about \(75\)\% of the blue-ray density.
		}\label{fig:fig6}
	\end{center}
\end{figure*}
%%%%%%%%%%%%%%%%%%%%%%%%%%%%%%%%%%%%%%%%%%%%%%%%%%%%%%%%%%%

\section{Towards higher information density}

In the last section of this article we want to assess if the information density of the individual nanostructures can be further increased.
As detailed in figure~\ref{fig:fig6}a, we therefore examine in the following the case of a geometry encoding 9 bit of information, which leads to 512 possible structural arrangements.
We note that the \(\approx 20\)\% larger size of the 9 bit structures compared to the 2-5 bit geometries leads to a red-shift of the resonances.
This effectively increases the information content in the red part of the spectra, which is advantageous for our aim to encode 9 bits per nanostructure.

\subsubsection{Training data and network architecture}

Having 512 different geometries to distinguish, the main difficulty is now the requirement of a large dataset.
Unfortunately, while no constraints exist concerning mass-production fabrication, on our equipment, which targets scientific work and maximum flexibility, acquiring scattering spectra from several hundreds of copies per geometry would imply many months of workload.
On the other hand, there are no general technical constraints and using dedicated, fully automated setups would enable a rapid experimental acquisition of very large datasets.
In our case however, in order to obtain a sufficiently large dataset, we numerically expand a small representative set of experimental spectra from 4 copies of each nanostructure.
For each nanostructure we generate 300 random superpositions of the 4 spectra, introducing additionally a random deviation of up to \(\pm 10\)\% in intensity.
The result of this numerical data expansion is shown at selected examples in figure~\ref{fig:fig6}b (see {\color{blue}SI}, Figs.~S26 and~S29-S32\comment{CHECK} for the whole experimental dataset and more examples).

\subsubsection{Results}

We train networks on 9 bit readout using either the full spectra or the scattering intensity at a discrete number of wavelengths (using simultaneously \(X\) and \(Y\) polarization).
The techniques are identical to the above cases of 2-5 bit per nanostructure (see also Methods).
In the left of figure~\ref{fig:fig6}c, the readout accuracy using full spectra is shown as a function of training samples per geometry. 
Quasi error-free operation is obtained for training-sets as small as about 20 spectra per digit.
Using a discrete number of probed wavelengths (Fig.~\ref{fig:fig6}c, right), very low error rates of \(\approx 2\)\% can be obtained already by probing three discrete wavelengths.
Using 4 or 5 probed wavelengths this can be improved to far below 1\%.

Finally, in order to assess how larger variations and noise in the data influence the readout performance of the ANN, we use a dataset of fully numerical simulations (using the Green Dyadic Method, see also Methods) of the 512 geometries.
This synthetic data allows us to adjust the distortions and noise of the spectra in a quantifiable way, which is shown schematically in the left of figure~\ref{fig:fig6}d.
The readout accuracy of an ANN, trained on full simulated spectra, is shown on the right of Fig.~\ref{fig:fig6}d.
Significant error rates start to occur only at noise levels around \(15\%\).
The average error rate with 20\% noise is still at a reasonable value of \(4.16\%\). 
For the worst digit though, the error rate is as high as \(35\%\) in this case, which however could be accounted for by designing optically ``more unique'' nanostructures for pathological cases.
Some examples of false and correct classifications as well as an analysis of the error rate for individual geometries at several noise levels is given in the {\color{blue}SI} Figs.~S34-S35.\comment{CHECK}
A comparison of the experimental data with simulations of different noise levels, a t-SNE plot of the experimental 9 bit dataset and further analysis on the impact of noise can be found in {\color{blue}SI} Fig.~S22 and~S24-S26.\comment{CHECK}

Let's finally compare the information density of the 9 bit nanostructures to the state-of-the art flat optical storage medium, the ``blue-ray disc''.
The blue-ray has a length and width per digit of \(150\,\)nm, respectively \(130\,\)nm, with \(320\,\)nm line spacing, leading to a required area \(A_{\text{1bit}} = (320+130) \cdot 150\,\text{nm}^2 = 67500\)\,nm\(^2\) per bit.
Our ``9 bit'' structures, require \((3\cdot 105 + 45 + \text{spacing})^2 / 9\,\)nm\(^2\) per bit.
In figure~\ref{fig:fig6}e we analyze the influence of a close neighbor on the scattering from a 9 bit nanostructure. 
Assuming a tolerable deviation of 10\% compared to the isolated nanostructure's spectrum, a spacing of \(300\)\,nm would still allow an accurate read-out, yielding around 40\% higher information density than the blue-ray (see also {\color{blue}SI} Fig.~S36-S37).\comment{CHECK}
With a \(300\)\,nm spacing, the 4 bit encoding nanostructures result in 75\% of the blue-ray information density.
% 2bit: 38\%
% 3bit: 56\%
% 4bit: 75\%
% 5bit: 84\%
% 9bit: 140\%
Using properly designed sets of photonic nanostructures with highly distinguishable optical responses, the accuracy and robustness of the method can be straightforwardly further optimized. 
Also, by increasing the structure height to red-shift the resonances, the area covered by each nanostructure could be reduced.
Finally, the information per structure could possibly be increased up to 11 or 12 bits (2048, respectively 4096 structure geometries). 
Larger binary sequences per structure seem however difficult for an accurate read-out, due to the power-scaling by which the number of structure geometries increases (\(2^{N_{\text{bit}}}\)).
A further means to improve the read-out accuracy and its robustness might be to include the angle of incidence as further probing parameter.
In order to exploit all three spatial dimensions of the storage medium, multi-layer arrangements might be possible.\cite{taylor_detuned_2012, taylor_electron-beam_2014}

%{\color{red}CONTINUE WORKING ON TEXT HERE \(\rightarrow\)}
%
%
%{\color{red}maybe add simulations to SI with higher structures to demonstrate red-shift}

\section{Conclusions}

In summary, we demonstrated on experimental data, that deep neural networks can be effectively trained for the optical retrieval of digital information, encoded in the geometry of photonic nanostructures.
We demonstrated on geometries encoding up to 9 bit per diffraction limited area, that the optical scattering spectra are more than sufficient for an accurate recovery of the encoded data. 
We showed that probing at a few discrete wavelengths or even simply using the RGB color information obtained from standard dark-field microscopy images, is a precise read-out scheme, potentially possible on very simple and cheap equipment.
The latter approach would also allow a massively parallel retrieval of the stored information.
Its robustness can easily be improved by employing nanostructure geometries with tailored, high color-contrast.\cite{gonzalez-alcalde_optimization_2018}

Our work paves the way towards accurate and massively parallel read-out of high density planar optical information using simple far-field characterization techniques combined with concepts of machine learning.
Re-writeable storage media might be created around the recently developed technology on catalytic magnesium metasurfaces for the dynamic adaptation of structural color.\cite{duan_dynamic_2017}
Our approach can be easily generalized to other classification tasks in nano-optics, including applications in plasmonics or in the identification of biological specimen.

\begin{acknowledgments}
	We gratefully thank Arnaud Arbouet and Christian Girard for their inspiring advise, their valuable help and for discussing and proof-reading the manuscript.
	This work was supported by Programme Investissements d'Avenir under the program ANR-11-IDEX-0002-02, reference ANR-10-LABX-0037-NEXT, by the LAAS-CNRS micro and nanotechnologies platform, a member of the French RENATECH network and by the computing facility center CALMIP of the University of Toulouse under grant~P12167.
\end{acknowledgments}

%% ------- contributions
\section*{Author contributions}
P.R.W. conceived the idea and designed the research together with G.L.
G.L. and A.L. developed the fabrication techniques. A.L. and N.M. fabricated the nanostructures and performed the electron microscopy.
P.R.W. carried out the optical experiments, did the simulations, the data analysis and implemented the machine learning part.
P.R.W. wrote the manuscript with contributions from G.L.
All authors discussed the results and commented on the manuscript at every stage.

%% ------- SUP. INFO.
\section*{Additional information}
Supplementary information is available in the online version of the paper. Reprints and permission information is available online at www.nature.com/reprints. Correspondence and requests for materials should be addressed to P.R.W.

%% ------- Competing interests
\section*{Competing financial interests}
The authors declare no competing financial interests.

%\bibliography{2018_multidigit_optical_storage}

\section{Methods}

\subsection{Nanofabrication of planar Si structures}

The silicon nanostructures were patterned on silicon on insulator (SOI) substrate (\(90\,\)nm active Si layer on \(30\,\)nm buried oxide) following a large scale top-down approach. 
Electron beam lithography was used to pattern \(80\,\)nm thick layer of an inorganic negative-tone resist, namely hydrogen silsesquioxane (HSQ). 
After exposure, HSQ was developed by immersion in 25\% tetramethylammonium hydroxide (TMAH) for one minute.\cite{guerfi_high_2013}
HSQ patterns were subsequently transferred to the silicon substrate down to the buried oxide by anisotropic reactive ion etching (RIE). 
Then, the structures were embedded in a \(200\,\)nm thick HSQ layer, deposited by spin coating in order to perfectly planarize the sample in a nanometrical range.\cite{guerfi_thin-dielectric-layer_2015} 
Finally, the HSQ layer (SiO\(_x\)H\(_y\)) was converted into a SiO\(_x\) layer by rapid thermal annealing at \(600^{\circ}\)C / 2 min under nitrogen ambiance, leading to a final SiO\(_x\) thickness layer of \(190\,\)nm.

%low-energy e-beam \cite{guerfi_high_2013}
%
%HSQ planarization \cite{guerfi_thin-dielectric-layer_2015}

\subsection{Confocal dark-field microscopy}

The scattering spectra were acquired out on a confocal dark-field microscope (Horiba XploRA).
A white lamp was focused on the sample by a \(\times 50\) dark-field objective (NA\,\(0.5\), condenser: NA\,\(0.8-0.95\)).
The backscattered light was filtered by a confocal hole (diameter of \(100\,\)\textmu m) and a polarization filter and dispersed by a grating (\(300\,\)grooves mm\(^{-1}\)) on a Peltier-cooled CCD.
The acquisition time was \(t_a = 0.2\)s.
All spectra were normalized by the spectrum of the lamp.
While the spectra used for the results shown in the main text were measured at an acquisition time of \(t_a = 0.2\,\)s, a second dataset was measured with \(t_a = 0.5\,\)s, leading to similar results (see {\color{blue}SI}, Figs.~S12-S14\comment{CHECK}). 
We therefore assume that even shorter acquisition times would be sufficient for a robust data recognition.

The polarization filtered dark-field images were taken using the same \(\times 50\) DF microscope objective, with a color CCD camera at a resolution of \(1392\times 1040\) and an exposure time of \(t_{\text{exp}}=0.05\,\)s. 
Each structure is perceived as a colored dot on the microscopy image, covering \(\approx 20\) -- \(50\) pixels, from which we take the average RGB value.

\subsection{Expansion of 9 bit experimental training data set}

Due to technical constraints of our measurement setup, we are not able to acquire in reasonable time, scattering spectra of several hundreds of all 512 possible 9 bit encoding nanostructures. 
In order to nevertheless assess the feasibility of experimental 9 bit readout via our machine learning based approach, we fabricate 4 copies of the 9 bit structures. 
To obtain the required much larger amount of spectra of each geometry, we subsequently create new spectra \(\sigma_{\text{new}}(\lambda)\) from random superpositions of the 4 experimental spectra \(\sigma_i(\lambda)\) of each nanostructure type.
Our condition in this approach is, that the random weights \(w_i\) of the four different spectra sum up to one, hence
\begin{equation}
	\sigma_{\text{new}}(\lambda) = \sum\limits_{i=1}^4 w_i \sigma_i(\lambda), \quad \text{with}\ \sum\limits_{i=1}^4 w_i = 1 \,.
\end{equation}
Finally, we multiply the spectrum by a random coefficient \(C\) between \(0.9\) and \(1.1\), to emulate larger intensity fluctuations.
In this way we generate 300 semi-experimental spectra with which we train the artificial neural networks for the decoding of the digital information.

\subsection{Electrodynamical simulations}

\subsubsection*{Simulated dataset}

The 9 bit geometry model used for the fully numerically simulated dataset is identical to the experimental structures (see figure~\ref{fig:fig6}a). 
It consists of \(3\times 3\) silicon blocks of each \(105\times 105\)\,nm\(^2\)  lateral size and \(90\,\)nm height.
Two sides are surrounded by an L-shaped block of \(45\)\,nm width. 

We numerically simulate the scattering spectra under \(X\) and \(Y\) polarized plane wave illumination for all 512 possible geometries using the Green Dyadic Method (GDM). 
In order to assess the robustness against different amount of perturbation and noise in the spectra, we numerically add noise to the simulated data via a sequence of random modifications. 
The noise addition steps are illustrated in the supporting informations, figure~S33. 
First we add random noise, apply a random positive offset as well as a
scaling factor. Then we multiply the spectra with a linear function of random slope and finally apply a wavelength-shift. In order to do so, in the first place we simulated the spectra on an extended wavelength range. 
Thanks to this procedure, we are capable of adjusting the magnitude of the random variations to yield more or less strongly distorted results. Figure~\ref{fig:fig6}d shows several randomized spectra for noise magnitudes between 10\% (i) and 25\% (iv).

\subsubsection*{Green Dyadic Method}

The numerical simulations for the 9-bit structures are performed using the Green Dyadic Method (GDM), a frequency-domain approach based on the concept of a generalized propagator.\cite{martin_generalized_1995}
In particular, we use an own implementation in python, ``pyGDM''.\cite{wiecha_pygdmpython_2018}

In the GDM the volume of a nanostructure is discretized with \(N\) cubic meshpoints of edge length~\(d\). 
To each of these mesh-points, a dipolar response is attributed. 
As detailed e.g. in reference\cite{girard_shaping_2008}, this leads to a system of \(3N\) coupled equations 
\begin{equation}\label{eq:inversion_equation}
\mathbf{E}_0 = \mathbf{M} \cdot \mathbf{E},
\end{equation}
which, by inversion of \(\mathbf{M}\), allows to relate any incident electric field \(\mathbf{E}_0\) to the induced field \(\mathbf{E}\) inside the particle.
\(\mathbf{M}\) is composed of \(3\times 3\) sub-matrices 
\begin{equation}\label{eq:submatrix_M}
\mathbf{M}_{ij} = \mathbf{I}\, \delta_{ij} - \alpha_i(\omega)\, \mathbf{G}(\mathbf{r}_i, \mathbf{r}_j, \omega).
\end{equation}
\(\mathbf{I}\) is the Cartesian unitary tensor, \(\delta_{ij}\) the Kronecker delta function and, in cgs units,
\begin{equation}
\alpha_i(\omega) = \frac{\epsilon_{i}(\omega) - \epsilon_{\text{env}}(\omega)}{4\pi} v_i\, .
\end{equation}
In the latter equation, \(v_i\) is the volume of the cubic discretization cells hence \(v_i = d^3\). 
We use the dispersion of silicon from Palik~\cite{palik_silicon_1997} for \(\epsilon_{i}\). 
The structures are placed in a homogeneous environment (\(\epsilon_{\text{env}}=1.45\) for SiO\(_2\)) at \(30\,\)nm above a silicon substrate.

\(\mathbf{G}\) in Eq.~\eqref{eq:submatrix_M} is the Green's Dyad, coupling the cubic meshpoints \(i\) and \(j\). 
It is composed of a vacuum term (accounting for the homogeneous environment) and a surface term (accounting for the substrate):
\begin{equation}\label{eq:GreensDyadFull}
\mathbf{G}(\mathbf{r}_i, \mathbf{r}_j, \omega) = 
\mathbf{G}_{0}(\mathbf{r}_i, \mathbf{r}_j, \omega)
+
\mathbf{G}_{\text{surf}}(\mathbf{r}_i, \mathbf{r}_j, \omega)
\end{equation}
which can be found in literature.\cite{girard_shaping_2008}
At \(\mathbf{r}_i=\mathbf{r}_j\) the Green's Dyad \(\mathbf{G}_{0}\) diverges, hence a normalization scheme is applied:
\begin{equation}
\mathbf{G}_{0}(\mathbf{r}_i, \mathbf{r}_i, \omega) = \mathbf{I}\, C(\omega)\, .
\end{equation}
For a cubic mesh we use\cite{girard_shaping_2008}  
\begin{equation}\label{eq:normalizationTerm}
C(\omega) = -\frac{4\pi}{3} \frac{1}{\epsilon_{\text{env}}(\omega) v_i} \, .
\end{equation}
We invert equation \eqref{eq:inversion_equation} using LU-decomposition,
the scattering cross-sections can be calculated from the near-field \(\mathbf{E}\) inside the particle.\cite{draine_discrete-dipole_1988}

\subsection{Training artificial neural networks for far-field scattering based classification}

The artificial neural networks are implemented in python using the tensorflow package.\cite{tensorflow2015-whitepaper}

\subsubsection*{Digit retrieval using scattering spectra -- 1D ConvNet}

\paragraph*{Preprocessing of scattering spectra}
Prior to the training of the ANN, we pre-process the acquired scattering spectra.
After background subtraction and normalization to the spectrum of the white lamp, we apply the following, further processing steps on our data.
We first apply a median filter with \(6\,\)nm kernel size to eliminate spikes from the spectra.
Subsequently we apply a smoothing filter based on moving averages with a \(15\,\)nm kernel to reduce the noise. 
We finally apply a down-sampling procedure using an order 8 Chebyshev type I filter by which we reduce the number of wavelengths to 99 per spectrum.

\paragraph*{Network architecture}
The scattering spectra based network for information retrieval is a one-dimensional convolutional network followed by a fully connected part.
The ConvNet consists of three layers using the ``leaky ReLU'' activation function. 
The first layer with 64 filters per channel uses a kernel of size 7.
The second and third layers both have 32 filters with kernels of size 5 and 3, respectively.
Each 1D convolutional layer is followed by a max pooling layer with kernel size 2 as well as by a batch normalization.\cite{ioffe_batch_2015}
The fully connected network consists of two layers with 64 and 32 neurons, employing a ``tanh'' activation.
For the ConvNet part, we apply an ``L2'' regularization, for the fully connected part a dropout regularization scheme with 80\% keep probability.
During training the output layer neuron whose index corresponds to the input binary number is set to ``1'' while the other neuron activations are kept at ``0''.
We train the network as depicted in Fig.~\ref{fig:fig1}d using a variant of the stochastic gradient descent (SGD) algorithm (``Adam'', for details see Ref.~\onlinecite{kingma_adam_2014}) with a batch size of 64.
We use the categorical cross-entropy loss, a learning rate of \(0.0001\) and train the network for \(200\) epochs. 
In the case of the 9 bit structures we train the network for \(2000\) epochs.

\subsubsection*{Discrete wavelengths and RGB digit retrieval -- Fully connected network}

\paragraph*{Discrete wavelengths: Selection and preprocessing}
The intensity values at specific wavelengths are extracted from the scattering spectra, acquired with white light illumination. We average the intensity over a small window of 3 neighboring values to reduce the impact of noise. Following spacings between wavelengths are used

\begin{itemize}
    \setlength\itemsep{0.em}
    \item first wavelength: \(\lambda_1=500\)\,nm
    \item 2 wavelengths: \(\lambda_2=630\)\,nm
    \item 3 wavelengths: \(\Delta \lambda = 95\)\,nm
    \item 4 wavelengths: \(\Delta \lambda = 75\)\,nm
    \item 5 wavelengths: \(\Delta \lambda = 60\)\,nm
\end{itemize}

Because we found that the red part of the spectra contains very little information (see also figure~\ref{fig:fig4}a), we set the longest wavelength to be not larger than about \(740\,\)nm in order to ideally probe the regions of the spectra, which contain the most information on the encoded bit sequence.

\paragraph*{RGB: Preprocessing of darkfield images}
We automatically process the polarization filtered DF microscopy images by considering all pixels with a brightness of at least \(3\times\) the background level. 
Using this procedure, each structure results in a dot of 30-50 pixels in the DF images. 
Our dataset is composed of the average RGB values of the ensemble of pixels corresponding to each nanostructure.
We normalize the RGB values to \(R+G+B=1\) and separately store the scattered intensity (hence the brightness).

\paragraph*{Network architecture}
For the RGB datasets as well as for the read-out using the scattering intensity at discrete wavelengths, we use a fully connected network.
The scattering intensity at the wavelengths \(\lambda_i\) or the R, G, B values (and optionally the intensity) are the input to the network, which itself consist of three layers with ``tanh'' activation in the following order: 128, 256 and 64 artificial neurons. 
We use L2 regularization and dropout with 80\% keep probability on the entire ANN.
The 16 neurons in the ``softmax'' output layer represent the binary numbers encoded in the nanostructures.
We train the network on data from 300 samples per geometry, using the original SDG algorithm and the categorical cross-entropy loss function.
The batch size is 64.
The learning rate is 1.0 with a \(\times 0.96\) decay each 1000 steps (\(\approx\) every 13 epochs).

\subsection{Considerations on the practical implementation of readout schemes and their performance}

\subsubsection*{Multi-wavelength based readout}
Using multiple lasers in a system similar to a 0.5em reader, it would be possible to probe the scattering intensity at different wavelengths. 
As in the case of 0.5em, the lasers could be scanning a rotating storage medium and acquire the spectral information in a sequential manner (for instance first a blue laser scans the structure, then a green and at last a red laser). 
The performance of such a system would be basically limited by the constraints of the available technology and should be comparable to the blue-ray disc, except that 9 bit of information (instead of a single bit) could be read at every passage of the laser system.

\subsubsection*{RGB based readout}
%% XPlora 50x objective, NA 0.5: 5.75 pixel per \textmu m
With our experimental setup, using a \(\times 50\), NA\,0.5 objective, we capture DF images covering areas of \(240\times 180\)\,\textmu m\(^2\). 
Assuming \(700\times 700\)\,nm\(^2\) per digit including spacing, we could capture around \(80,000\) structures per image. In the case of 5 encoded bit per geometry, these nanostructures would encode approximately \(0.44\)\,Mbit of information.
The 0.5em disc at its \(1\times\) datarate yields \(36\)\,Mbit/s, so around 100 images per second would be necessary to yield similar performance with our approach.
Throughout this study, we used an exposure time of \(t_{\text{exp}}=0.05\,\)s, which, under otherwise perfect conditions, would correspond to 20 images per second. 
This is still about 5 times below 0.5em performance.

Our setup however is far from ideal for this specific readout task. 
Using high-NA and low magnification objectives, the captured area could easily be significantly enlarged. 
For instance a \(\times 20\), NA\,0.75 microscope objective would cover a \(2.5\times 2.5\) times larger area than our microscope, which would lead to \(2.75\,\)Mbit per image (again using the 5 bit structures).
Also the exposure time could easily be strongly decreased using very bright light sources and optimized CCD cameras.
Assuming \(t_{\text{exp}}=0.01\)s (which is still a very conservative guess), this would then enable a readout rate of \(275\)MBit/s.

\subsection{Data availability}
The authors declare that all data supporting the findings of this study are available within the article and its Supplementary Information files or from the corresponding author upon reasonable request.

\subsection{Code availability}
The authors declare that all software used to obtain the results of this work are publicly accessible as open-source software: \href{https://www.python.org/}{python} including \href{https://www.scipy.org/}{scipy}, \href{https://www.tensorflow.org/}{tensorflow}, as well as \href{https://wiechapeter.gitlab.io/pyGDM2-doc/}{pyGDM}\cite{wiecha_pygdmpython_2018}, our own implementation of the Green Dyadic Method.
Our scripts can be made accessible from the corresponding author upon reasonable request.

\end{document}